\documentclass[12pt,thmsa]{article}
\usepackage{sw20lart}

\input tcilatex2.5
\QQQ{Language}{
British English
}

\begin{document}

\title{Varying G and Other Constants}
\author{John D. Barrow \\
Astronomy Centre\\
University of Sussex\\
Brighton BN1 9QJ\\
UK}
\maketitle

\begin{abstract}
We review recent progress in the study of varying constants and attempts to
explain the observed values of the fundamental physical constants. We
describe the variation of $G$ in Newtonian and relativistic scalar-tensor
gravity theories. We highlight the behaviour of the isotropic Friedmann
solutions and consider some striking features of primordial black hole
formation and evaporation if $G$ varies. We discuss attempts to explain the
values of the constants and show how we can incorporate the simultaneous
variations of several 'constants' exactly by using higher-dimensional
unified theories. Finally, we describe some new observational limits on
possible space or time variations of the fine structure constant.
\end{abstract}

\section{Introduction}

In this overview of some aspects of varying constants we will begin by
considering the time variation of the gravitational 'constant' $G$ in
Newtonian and relativistic theories of gravity. Although the Newtonian
situation is usually ignored, it provides a number of instructive parallels
and contrasts with the more complex situation that prevails in scalar-tensor
generalisations of general relativity. We will focus upon the behaviour of
the cosmological solutions in these theories and provide prescriptions for
generating the isotropic Friedmann solutions to any version of a
scalar-tensor gravity theory.

Next, we shall highlight the unusual situation that seems to be created if a
primordial black hole forms in the very early stages of a universe in which $%
G$ is changing with time. Then we shall go on to consider some of the
speculative ideas that have been put forward to explain the values of the
constants of Nature. We shall discuss the problem of the simultaneous
variation of several 'constants' and describe how this situation can be
modelled using simple scaling invariances of physics or by exploiting the
structure of unified higher-dimensional theories of the fundamental
interactions. One of the most interesting quantities to appear in these
discussions is $\alpha ,$ the fine structure constant. In the final section,
we describe some new observational limits on any possible space or time
variations in the fine structure constant that can be deduced from
spectroscopic observation of molecular and atomic hydrogen absorption lines
from the gas around radio-loud quasars. 

\section{Some Background to Varying G}

The study of gravitation theories in which Newton's gravitational constant
varies in space and time has many motivations. It began in 1935 with the
proposal by Milne of a theory of gravitation with two time standards (one
for gravitational processes, the other for atomic processes) in which the
mass within the particle horizon, M$_h\propto c^3G^{-1}t,$ remains constant
with respect to $t$-time, led to the prediction that $G\propto t$ in this
time. The idea became of wider interest in 1937 with the 'Large Numbers
Hypothesis' of Dirac (1937a,b 1938), that the ubiquity of certain large
dimensionless numbers, with magnitudes $O(10^{39})$, which were known to
arise in combinations of physical constants and cosmological quantities
(Weyl 1919, Zwicky 1939, Eddington 1923) was not a coincidence but a
consequence of an underlying relationship between them (Barrow and Tipler
1986, Barrow 1990a). This relationship required a linear time variation to
occur in the combination $e^2G^{-1}m_N$ (where $e$ is the electron charge, $%
m_N$ the proton mass, and $G$ the Newtonian gravitation constant) and Dirac
proposed that it was carried by $G\propto t^{-1}$, (Chandrasekhar 1937,
Kothari 1938)$\ $This led to a range of new geological and palaeontological
arguments being brought to bear on gravitation theories and cosmological
models (Jordan 1938, 1952, Teller 1948, Dicke 1957, 1964, Gamow 1967a,b).
Brans and Dicke (1961) refined the scalar-tensor theories of gravity first
formulated by Jordan and, motivated by apparent discrepancies between
observations and the weak-field predictions of general relativity in the
solar system, proposed a generalization of general relativity that became
known as Brans-Dicke theory. As the solar system and binary pulsar
observations have come into close accord with the predictions of general
relativity so the scope for a theory of Brans-Dicke type to make a
significant difference to general relativity in other contexts, notably the
cosmological, has been squeezed into the very early universe. However, more
general theories with varying $G$ exist, in which the Brans-Dicke parameter
is no longer constant (Barrow 1993a). These theories possess cosmological
solutions which are compatible with solar-system gravitation tests (Hellings
1984, Reasenberg 1983, Shapiro 1990, Will 1993), gravitational lensing
(Krauss and White 1992), and the constraints from white-dwarf cooling (Vila
1976, Garc\'ia-Berro et al 1995). The crucial role that scalar fields may
have played in the very early universe has been highlighted by the
inflationary universe picture of its evolution. A scalar field, $\phi $,
which acts as the source of the gravitational coupling, $G\sim \phi ^{-1}$,
is a possible source for inflation and would modify the form of any
inflation that occurs as a result of the universe containing weakly coupled
self-interacting scalar fields of particle physics origin. There have been
brief periods when experimental evidence was claimed to exist for a
non-Newtonian variation in the Newtonian inverse-square law of gravitation
at low energies over laboratory dimensions (Fischbach et al 1986) and
speculations that non-Newtonian gravitational behaviour in the weak-field
limit might explain the flatness of galaxy rotation curves (Bekenstein and
Meisels 1980, Milgrom 1983, Bekenstein and Milgrom 1984, Bekenstein and
Sanders 1994) usually cited as evidence for the existence of non-luminous
gravitating matter in the Universe. Most recently, particle physicists have
discovered that space-times with more than four dimensions have special
mathematical properties which make them compelling arenas for
self-consistent, finite, anomaly-free, fully-unified theories of four
fundamental forces of Nature (Green and Schwarz 1984). Our observation of
only three large dimensions of space means that some dimensional segregation
must have occurred in the early moments of the universal expansion with the
result that all but three dimensions of space became static and confined to
very small dimensions $\sim 10^{-33}cm.$ Any time evolution in the mean size
of any extra ($>3$) space dimensions will be manifested as a time evolution
in the observed three-dimensional coupling constants (Freund 1982, Marciano
1984, Kolb, Perry and Walker 1986, Barrow 1987). The effect of this
dimensional reduction process is to create a scalar-tensor gravity theory in
which the mean size of the extra dimensional behaves like a scalar field. In
particular, low-energy bosonic superstring theory bears a close relation to
a particular limit of Brans-Dicke theory (see section 4).

However, despite these interconnections with modern ideas in the cosmology
of the early universe, the theoretical investigation of gravity theories
with time-varying $G$ is still far from complete and, aside from the solar
system and binary pulsar observations (Will 1993), there are few general
observational restrictions on scalar-tensor theories which are clear-cut.

\section{Newtonian Varying G\textbf{\ }}

We shall begin by investigate Newtonian gravity theories with varying $G$,
pointing out the relationships that these simple solutions have to the more
complicated solutions of scalar-tensor gravity theories. In the past there
has been very little discussion of the Newtonian case (see Barrow 1996). The
exceptions are the rediscoveries of Meshcherskii's theorem (1893, 1949): for
example, by Batyrev (1941,1949), Vinti (1974), Savedoff and Vila (1964),
Duval, Gibbons and Horv\'athy (1991), McVittie (1978) and Lynden Bell
(1982). These authors all recognised the equivalence of the Newtonian
gravitational problem with time-varying $G$ to the problem with constant $G$
and varying masses.

Newtonian gravitation is a potential theory that is derived from the axiom
that the external gravitational potential due to a sphere of mass M be equal
to that of a point of mass M. This fixes the potential to be equal to

\begin{equation}
\Phi (r)=\frac Ar+Br^2
\end{equation}
where A and B are constants; $A=-GM$ and $B=\frac 16\Lambda $, where $%
\Lambda $ is the cosmological constant of Einstein. This argument shows how
the cosmological constant arise naturally in Newtonian theory, as it does in
general relativity. Unless otherwise stated, we shall set the cosmological
constant term zero ($B=0=\Lambda )$. In section 3.2 we shall discuss how its
interpretation differs from a $p=-\rho $ fluid when $G$ is not constant and
prove some restricted cosmic no hair theorems.

Consider the Newtonian N-body problem with a time-varying gravitational
'constant' $G(t)$. If the N bodies have masses m$_j$ and position vectors $%
\mathbf{r}_j$ then

\begin{equation}
\frac{d^2\mathbf{r}_j}{dt^2}=-\sum G(t)m_k\frac{\mathbf{r}_j-\mathbf{r}_k}{%
\left| \mathbf{r}_j-\mathbf{r}_k\right| ^3}.
\end{equation}
Now, if we have a solution, $\mathbf{\hat r}_j(\hat t),$ of these equations
with $G=G_0$ independent of time, then,

\begin{equation}
\mathbf{r}_j(t)=(\frac{t+b}{t_0})\mathbf{\hat r}_j(-\frac{t_o^2}{t+b}+c)
\end{equation}
is an exact solution of the equations (1) with

\begin{equation}
G(t)=G_0\times \left( \frac{t_0}{t-c}\right)
\end{equation}
where $b,c$ and $t_0$ are constants with $t_0\neq 0.\ $Thus, given any
solution of a gravitational problem (for example the output from a
cosmological N-body gravitational clustering simulation) with constant $G$
we can immediately write down an exact solution in which $G$ varies
inversely with time. For example, suppose we take the simplest Newtonian
cosmological model with zero total energy, when $G=G_o$ is constant. Then,
the expansion scale factor of the universe is

\begin{equation}
\hat r\propto \hat t^{2/3}.
\end{equation}
By the theorem we have that

\begin{equation}
r(t)=(\frac{t+b}{t_0})(-\frac{t_o^2}{t+b}+c)^{2/3}
\end{equation}
when $G(t)$ varies as

\begin{equation}
G(t)=G_0\times \left( \frac{t_0}{t-c}\right) \ ;t_0\neq 0
\end{equation}
and so $r(t)\propto t^{1/3}$ as $t\rightarrow \infty .$

The result (4) is also useful for modelling small variations in $G$ over
short timescales. If we expand an arbitrary analytic form for $G(t)$ to
first order in $t$ then

\begin{equation}
G(t)=G_0+\dot G_0t+....O(t^2)\approx G_0(1-t\dot G_0/G_0)^{-1}
\end{equation}
and this has the form (4).

This result, a consequence of the scale invariance of the inverse-square law
of force, was first found by Meshcherskii (1893). It has often been
rediscovered and elaborated. Duval, Gibbons and Horv\'athy (1991) have
explored its existence in a wider context and displayed similar invariances
of the non-relativistic time-dependent Schr\"odinger equation with Coulomb
potential (see also Barrow and Tipler 1986) which enables solutions with
time-varying electron charge ($e^2\propto t)$ to be generated by
transformation of known exact solutions with constant values of $e$. In the
next section we shall prove a generalization of Meshcherskii's theorem for
cases where the pressure is non-zero and the equation of state has perfect
fluid form. 

\subsection{Newtonian Cosmologies with $\mathbf{G(t)\propto t}^{-n}$}

We adopt the standard generalization of Newtonian cosmology (Milne and
McCrea 1934, Heckmann and Sch\"ucking 1955, 1959) to include matter with
non-zero pressure and a perfect fluid equation of state. We shall confine
our attention to isotropic Newtonian solutions. This is of particular
interest for the real universe in the recent past but we also know that
anisotropic Newtonian cosmological models are not well posed in the sense
that there are insufficient Newtonian field equations to fix the evolution
of all the degrees of freedom (there are no propagation equations for the
shear anisotropies (Barrow and G\"otz 1989a)) and this incompleteness must
be repaired by supplementing the theory with extra boundary conditions or by
importing shear propagation equations from a complete relativistic theory,
like general relativity (Evans 1974, 1978, Shikin 1971, 1972), or by
ignoring the evolution of the shear anisotropy (Narlikar 1963, Narlikar and
Kembhavi 1980, Davidson and Evans 1973, 1977).

Consider a homogeneous and isotropic universe with expansion scale factor $%
r(t).\ $The material content of the universe is a perfect fluid with
pressure, $p$, and density $\rho ,$ obeying an equation of state (where the
velocity of light has been set equal to unity)

\begin{equation}
p=(\gamma -1)\rho ;\text{ }0\leq \gamma \leq 2,
\end{equation}
with $\gamma $\ constant. If $G=G(t)$ then the equation of motion for $r(t)$
is

\begin{equation}
\ddot r(t)=-\frac{G(t)M}{r^2}=-\frac{4\pi G(t)(\rho +3p)r}3.
\end{equation}
The mass conservation equation is

\begin{equation}
\dot \rho +3\frac{\dot r}r(\rho +p)=0.
\end{equation}
Hence, we have

\begin{equation}
\rho =\frac \Gamma {r^{3\gamma }};\text{ }\Gamma \geq 0,\text{ }constant.
\end{equation}
We shall initially be interested in power-law variations of $G(t)$ of the
form

\begin{equation}
G(t)=G_0\left( \frac{t_0}t\right) ^n
\end{equation}
so we have

\begin{equation}
\ddot r=-\lambda t^{-n}r^{1-3\gamma }
\end{equation}
where $\lambda $ is a constant defined by

\begin{equation}
\lambda =\frac{4\pi G_0t_0^n(3\gamma -2)\Gamma }3
\end{equation}
so the sign of $\lambda $ is determined by the sign of $3\gamma -2$, as in
isotropic general relativistic cosmologies. Hence, accelerating universes ($%
\ddot r>0)$ arise when $3\gamma >2$ regardless of whether $G$ varies or not.
However, these accelerating universes need not solve the horizon and
flatness problems in the way that conventional inflationary universes do;
that depends upon the value of $n$.

A generalization of Meshcherskii's theorem can be proved for the case with $%
p=(\gamma -1)\rho .$ If $\hat r(\hat t)\ $is a solution with $G=G_0$
constant, then (Barrow 1996),

\begin{equation}
r(t)=(\frac{t+b}{t_0})\hat r(-\frac{t_o^2}{t+b}+c)\ 
\end{equation}
with $b,c,$ and $t_0\neq 0,$ constants, is an exact solution of (14) with

\begin{equation}
G(t)=G_0\times \left( \frac{t_0}{t-c}\right) ^{4-3\gamma };t_0\neq 0.
\end{equation}
These results provide a Newtonian analogue to the conformal properties of
relativistic scalar-tensor theories. We can draw a number of general
conclusions from them. As $t\rightarrow \infty $ we have

\begin{equation}
r(t)\rightarrow t,\text{ if }c\neq 0,\text{ }\forall \gamma
\end{equation}

\begin{equation}
r(t)\rightarrow \frac t{t_0}\hat r\left( \frac{-t_{0^{}}^2}t\right)
^{2/3\gamma },\text{ if }c=0\text{ and }\gamma \neq 0.
\end{equation}
In particular, if we take the solutions with constant $G=G_0$ to be the
zero-curvature Friedmann solutions then, when $c\neq 0,$ we have

\begin{equation}
\hat r(\hat t)\propto \hat t^{2/3\gamma },\text{ if }\gamma \neq 0\ 
\end{equation}

\begin{equation}
\hat r(\hat t)\propto \exp [H_0\hat t],\text{ }H_0\text{ constant, if }%
\gamma =0,\ 
\end{equation}
and the solutions with $G(t)\propto t^{3\gamma -4}$ at large time have the
form

\begin{eqnarray}
r(t)\ &\propto &t^{(3\gamma -2)/3\gamma \text{ }},\text{ if }\gamma \neq
0\neq c  \nonumber \\
r(t) &\propto &t,\text{ if }\gamma \neq 0,\text{ }c=0
\end{eqnarray}

\begin{equation}
r(t)\propto t\exp \left[ H_0(c-\frac{t_0^2}t)\right] \ \rightarrow t,\text{
if }\gamma =0,\forall c.
\end{equation}
These are particular solutions only, of course, and their properties need
not be shared by the general solutions for a given value of $n$ or $\gamma $%
. The $\gamma =0$ solution, (23), does not exhibit inflation and is
asymptotic to the solution of the equation $\ddot r=0$. This is a result of
the very rapid decay of $G(t)\propto t^{-4}.$

When $\gamma =4/3$ there is no possible time-variation of $G$ which
preserves the scaling invariance and for other positive values of $\gamma $
the expansion is slower than in universes with constant $G;\ $power-law
inflation does not occur in the varying-$G$ solutions when $0<\gamma <2/3.$

Equation (14) describes motion under a time-dependent force for which there
need exist no time-independent energy integral. Therefore we cannot write
down a Friedmann equation for $\dot r$ in the usual way. However, there
exists a class of particular exact solution with simple power-law form:

\begin{equation}
r(t)\propto t^{(2-n)/3\gamma };\text{ }\gamma \neq 0
\end{equation}
\begin{equation}
\rho (t)=\frac{(2-n)(3\gamma -2+n)}{12\pi G_0\gamma ^2t_0^n(3\gamma
-2)t^{2-n}}=\frac{(2-n)(3\gamma -2+n)}{12\pi \gamma ^2(3\gamma -2)G(t)t^{2\ }%
}
\end{equation}
so $\rho \geq 0$ requires that

\begin{equation}
\frac{(2-n)(3\gamma -2+n)}{(3\gamma -2)^{\ }}\geq 0.
\end{equation}
Clearly, when $n=0$, these solutions reduce to the familiar zero-curvature
(zero energy) solutions of general relativistic (Newtonian) cosmology with
constant $G$. They describe expanding universes so long as $n<2.\ $They are
particular solutions because they do not possess the full complement of
arbitrary constants of integration that specify the general solution.
Solutions of this sort suggest that there may exist more general solutions
that behave at early times like a solution of the form (24) with one value
of $n=n_1$ for $t\leq t_1$ and with another value $n=n_2$ for $t\geq t_1.$ A
'bouncing' solution would have $n_1>2$ and $n_2<2$, for example, with $%
t_1<<t_0$. There are many examples of scalar-tensor gravity theories with
cosmological models that display this early-time behaviour (Barrow 1993b,
Barrow and Parsons 1996).

We shall not go on to discuss the general solutions of the Newtonian
evolution equation and the circumstances under which they approach these
particular solutions. Such a analysis can be found in Barrow (1996).

\subsection{Inflationary universe models with $\mathbf{p=-\rho }$}

Scalar-tensor gravity theories have provided an arena in which to explore
variants of the inflationary universe theory first proposed by Guth (1981)
in which inflation is driven by the slow evolution of some weakly-coupled
scalar field. The scalar field from which the gravitational coupling is
derived can in principle be the scalar field from which the gravitational
coupling is derived or it can influence the form of inflation produced by
some other explicit scalar matter field. A number of studies have been made
of the behaviour of inflation in scalar-tensor gravity theories
(Mathiazhagen and Johri 1984, La and Steinhardt 1989, Barrow and Maeda 1990,
Steinhardt and Accetta 1990, Garc\'ia-Bellido, Linde and Linde 1994, Barrow
and Mimoso 1994, Barrow 1995). The non-linear master equation governing the
evolution for $r(t)\ $has interesting behaviour in the inflationary cases
where $\ddot r>0.$

The particular power-law solutions (24)-(26) with $2\geq \gamma >0$ expand
when $n<2.$ Although they accelerate with time ($\ddot r>0$) whenever $%
3\gamma -2<0,$ the expansion only provides a possible solution of the
horizon problem when

\begin{equation}
2-n>3\gamma >0.
\end{equation}
So, in the case of radiation $(\gamma =4/3)$ the horizon problem can be
solved if $n<-2\ $in the early stages of the expansion.

In the most interesting case, when $\gamma =0,$ and $\rho $ is constant, the
perfect-fluid matter source mimics the behaviour of a slowly rolling scalar
field whose evolution is dominated by its self-interaction potential. When $%
\gamma =0$ the master evolution equation (14) is linear in $r$

\begin{equation}
\ddot r=-\lambda t^{-n}r
\end{equation}
with $\lambda <0.$ We are interested in determining the asymptotic behaviour
of this equation as $t\rightarrow \infty $ for all values of $n$ in order to
determine when there is asymptotic approach to the usual de Sitter solution
that obtains when $n=0.$ The solutions fall into three classes according to
the value of $n.$ For $n<2$, the solutions asymptote towards the WKB
approximation as $t\rightarrow \infty $

\begin{equation}
r(t)\sim t^{n/4}\exp \left\{ \frac{2H_0t^{\frac{2-n}2}}{2-n}\right\} \text{ }%
;\text{ }n<2
\end{equation}
where the constant Hubble parameter, $H_0,$ is given by

\begin{equation}
H_0^2\equiv -\lambda =-\frac{4\pi G_0t_0^n(3\gamma -2)\Gamma }3
\end{equation}
which is positive for $3\gamma -2<0.$

For $n>2$ the asymptote is

\begin{equation}
r(t)\sim t\text{ ; }n>2.
\end{equation}
In fact, this is a particular case of a stronger result that does not assume
that $G(t)$ is a power-law. The asymptote $r\sim t$ results whenever $G(t)$
falls fast enough to satisfy (Cesari 1963)

\begin{equation}
\int\limits^\infty t\mid G(t)\mid dt<\infty .
\end{equation}

When $n=2$ the asymptote is

\begin{eqnarray}
r(t) &\sim &t^\alpha \\
\alpha &=&\frac 12(1+\sqrt{1+4A})  \nonumber \\
A &\equiv &\frac{8\pi G_0\rho }{3t_0^n}>0.  \nonumber
\end{eqnarray}

Thus we see that if $G(t)$ falls off faster than $t^{-2}$ the solutions of
eqn. (34) approach those of $\ddot r=0$ and no inflation occurs. By
contrast, if $G(t)$ falls off more slowly that $t^{-2}$, grows ($n<0),$ or
remains unchanged ($n=0),$ then inflationary solutions of the form (29)
arise.$\ $We notice that in the absence of $G$-variation $(n=0)$ this
solution reduces to the well known de Sitter expansion $(r(t)\propto \exp
\left\{ H_0t\right\} )$ familiar in general relativistic models of inflation
with a constant vacuum energy density or positive cosmological constant.
When $0<n<2$ it produces a form of sub-exponential inflation which is
familiar from studies of scalar-tensor gravity theories with varying $G(t)$
and models of intermediate inflation studied in general relativistic
cosmologies containing a wide range of scalar fields, (Barrow 1990b, Barrow
and Saich 1990, Barrow and Liddle 1993). If $n<0$ there is super-exponential
inflation.

There is a Newtonian 'no hair' theorem in the general case where no
particular form is assumed for the time-variation of $G(t)\ $because we can
make use of the general asymptotic properties of the evolution equation. We
have already given this result for cases where $G(t)$ falls off faster than $%
t^{-2}$ as $t\rightarrow \infty $ in eqns. (31)-(32), and as $t^{-2}$ in
eqn. (33)$.$ In the case where the fall-off is slower than $t^{-2}$ we have
a WKB approximation

\begin{equation}
r(t)\sim c\left[ G(t)\right] ^{-\frac 14}\exp \left\{ \omega \int \left[
G(t)\right] ^{\frac 12}dt\right\}
\end{equation}
as $t\rightarrow \infty ,$ where $\omega ^3=1$ and $c$ is a constant, so
long as

\begin{equation}
\left| t^2G(t)\right| \rightarrow \infty \text{ as }t\rightarrow \infty .
\end{equation}
Clearly, eqn. (29) gives this asymptote in the special case that $\
G(t)\propto t^{-n}$ with $n<2$ when we choose the $\omega =1$ part of the
linear combination of solutions$.$

When $G$ is constant in general relativity and in Newtonian gravitation the
presence of a perfect fluid with an equation of state $p=-\rho $ is
equivalent to the addition of a cosmological constant to the field
equations. However, in Newtonian gravitation with varying $G$ and in
scalar-tensor generalizations of general relativity, these two are no longer
equivalent. If we had included a cosmological constant term in the Newtonian
gravitational equation (10), it would become

\begin{equation}
\ddot r(t)\ =-\frac{4\pi G(t)(\rho +3p)r}3+\frac{\Lambda r}3
\end{equation}
and we can see that the choice $p=-\rho =$ constant, using (11), only makes
the first term on the right-hand side of (36) equivalent to the term $%
\propto \Lambda r$ if $G(t)$ is constant. Clearly, the late-time may not be
dominated by the cosmological constant term when $G(t)\propto t^{-n}$ and $%
n<0.$ We will not investigate the behaviour for general $\gamma $ here, see
Barrow (1996). 

\section{Relativistic scalar-tensor theories}

Newtonian gravitation permits us to 'write in' an explicit time variation of 
$G$ without the need to satisfy any further constraint. However, in general
relativity the geometrical structure of space-time is determined by the
sources of mass-energy it contains and so there are further constraints to
be satisfied. Suppose that we take Einstein's equation in the form ($c\equiv
1$)

\begin{equation}
\tilde G_b^a\ =8\pi GT_b^a
\end{equation}
where $\tilde G_b^a$ and $T_b^a$ are the Einstein and energy-momentum
tensors, as usual, but imagine that $G=G(t)$. If we take a covariant
divergence $\ _{;a}$ of this equation the left-hand side vanishes because of
the Bianchi identities, $T_{b;a}^a=0$ if energy-momentum conservation is
assumed to hold, hence $\ \partial G/\partial x^a=0\ $ always. In order to
introduce a time-variation of $G$ we need to derive the space and time
variations from some scalar field, $\psi $ which then contributes a stress
tensor $\tilde T_b^a(\psi )$ to the right-hand side of the gravitational
field equations.

We can express this structure by a choice of lagrangian, linear in the
curvature scalar $R,$ that generalizes the Einstein-Hilbert lagrangian of
general relativity with,

\begin{equation}
L=-f(\psi )R+\frac 12\partial _a\psi \partial ^a\psi +16\pi L_m
\end{equation}
where $L_m$ is the lagrangian of the matter fields. The choice of the
function $f(\psi )$ defines the theory. When $\psi $ is constant this
reduces, after rescaling of coordinates, to the Einstein-Hilbert lagrangian
of general relativity. (We ignore, for simplicity the possibility of
including a cosmological 'constant' term which can now be a function of the
scalar field $\psi ).$ For historical reasons scalar-tensor theories have
not been written with a gravitational lagrangian of this simple form
(Bergmann 1968, Steinhardt and Accetta 1990, Holman et al 1991) but have
followed the formulation introduced by Brans and Dicke (1961). This can be
obtained from (38) by a non-linear transformation of $\psi $ and $f$. Define
a new scalar field $\phi ,$ and a new coupling function $\omega (\phi )$ by

\begin{eqnarray}
\phi &=&f(\psi ) \\
\omega (\phi ) &=&\frac f{2f^{\prime 2}}  \nonumber
\end{eqnarray}
then (75) becomes

\begin{equation}
L=-\phi R+\frac{\omega (\phi )}\phi \partial _a\phi \partial ^a\phi +16\pi
L_m.
\end{equation}
This has the Brans-Dicke form. The Brans-Dicke theory arises as the special
case

\begin{equation}
Brans-Dicke:\text{ }\omega (\phi )=\text{ constant; }f(\psi )\propto \psi ^2.
\end{equation}
However, in general, there exists an infinite number of these theories
defined by the choice of $\omega (\phi ).$ One of the reasons for renewed
interest in scalar-tensor gravity theories of this form is the relationship
that exists between the gravitational part of lagrangian (i.e. excluding $%
L_m $) for Brans-Dicke theory and the low-energy effective action for
bosonic string theory, which can be written as (Callan et al 1985)

\begin{equation}
L_{sst}=\exp (-2\chi )(R+4\chi ^{,a}\chi _{,a}-\frac 1{12}H^2)
\end{equation}
where $\chi $ is the dilaton field and $H^2\equiv H_{abc}H^{abc}$, where $%
H_{abc}$ is the totally antisymmetric 3-form field. If we identify $\phi
=\exp (-2\chi )$ then $L_{sst}$ is identical to the Brans-Dicke lagrangian,
(40), when $\omega =-1.$ However, differences arise in the couplings of the
scalar fields to other forms of matter in the two theories.

$\ $ The field equations that arise by varying the action associated with $L$
in (40) with respect to the metric, $g_{ab},$ and $\phi $ separately, gives
the field equations,

\begin{eqnarray}
\tilde G_{ab} &=&\frac{-8\pi }\phi T_{ab}-\frac{\omega (\phi )}{\phi ^2}%
\left[ \phi _{,a}\phi _{,b}-\frac 12g_{ab}\phi _{,i}\phi ^{,i}\right] \\
&&-\phi ^{-1}\left[ \phi _{,a;b}-g_{ab}\Box \phi \right]  \nonumber
\end{eqnarray}
\begin{equation}
\left[ 3+2\omega (\phi )\right] \Box \phi =8\pi T_a^a-\omega ^{\prime }(\phi
)\phi _{,i}\phi ^{,i}
\end{equation}
where the energy-momentum tensor of the matter sources obeys the
conservation equation

\begin{equation}
T_{;a}^{ab}=0.
\end{equation}
These field equations reduce to those of general relativity when $\phi $
(and hence $\omega (\phi ))$ is constant, in which case the Newtonian
gravitational constant is defined by $G=\phi ^{-1}.$ An interesting feature
of these equations is clear by inspection: when the trace of the
energy-momentum tensor vanishes (this includes vacuum and
radiation-dominated solutions as important particular cases) any solution of
general relativity is a particular solution of the scalar-tensor theory with 
$\phi $ (and hence $G$) constant.

These equations are also conformally related to general relativity when the
trace $T_a^a=0$. If we conformally transform the metric

\begin{equation}
g_{ab}\rightarrow \Omega ^{-2}g_{ab}
\end{equation}
with $\Omega =\phi ^{1/2}$ and define $\Psi $ by

\begin{equation}
\Psi =\int \sqrt{\frac{2\omega (\phi )+3}2}\frac{d\phi }\phi
\end{equation}
then the conformally transformed theory is general relativity with a matter
source consisting of a scalar field $\Psi $ with a potential $V(\Psi ).$
This conformal invariance can be exploited to produce powerful
solution-generating procedures for scalar-tensor theories (Barrow 1993a,
Barrow and Mimoso 1994, Barrow and Parsons 1996). One can regard
Meshcherskii's theorem and its generalization proved above in (16)-(17) as
Newtonian analogues of these conformal invariances.

Although the coupling function $\omega (\phi )$ is unconstrained by the
structure of scalar-tensor gravity theories, its choice determines the form
of the cosmological models in the theory and the form of the weak-field
limit. Typically, the weak-field solar-system predictions of scalar tensor
gravity theories have the following relationship to those of general
relativity (GR)

\begin{equation}
\omega (\phi )\text{ weak field result}\approx \text{ (GR result)}\times
\left[ 1+O\left( \frac{\omega ^{\prime }}{\omega ^3}\right) \right] .
\end{equation}

Specifically, the perihelion precession of Mercury, $\Delta \vartheta ,$ and
the time-variation of $G$ is predicted at second-order to be (Nordvedt 1970,
Wagoner 1970, Will 1993)

\begin{eqnarray}
\Delta \vartheta &\simeq &43^{\prime \prime }\times \left[ 1-\frac
1{12+6\omega }\left\{ 4+\frac{\omega ^{\prime }}{(3+2\omega )^2}\right\}
\right] \text{ per 100 yr} \\
\frac{\dot G}G &=&-\left( \frac{3+2\omega }{4+3\omega }\right) \left[ \frac{%
G(t)}{G_0}+\frac{2\omega ^{\prime }(\phi )}{(3+2\omega )^2}\right] \dot \phi
\end{eqnarray}
where $G_0$ is the present measured value of $G(t)$. There is even a special
theory for which the term in $\left[ ...\right] $ brackets vanishes in (50)
and $G(t)=G_0$ is constant to this weak-field order (Barker 1978). In
Brans-Dicke theory we have $G\propto \phi ^{-1}.$ The general relativistic
limit of an $\omega (\phi )$ scalar-tensor theory is obtained (if it exists)
by taking the two limits

\begin{equation}
\omega \rightarrow \infty \text{ and }\frac{\omega ^{\prime }}{\omega ^3}%
\rightarrow 0.
\end{equation}
The observational limits on $\omega $ require $\omega >500$ but the limit on 
$\omega ^{\prime }$ is weak, $\omega ^{\prime }<O(1)$, (see Will 1993)$.\ $%
Similar limits are obtained from the binary pulsar (Damour, Gibbons \&
Tayler 1988), but are more model dependent.

\subsection{Scalar-tensor Friedmann cosmologies}

In order to examine some of the counterparts to the Newtonian solutions
discussed above we give the field equations for an isotropic and homogeneous
Friedmann universe with scale factor $a(t),$ curvature parameter $k$, and
Hubble rate $H=\dot a/a$,

\begin{equation}
H^2=\frac{8\pi \rho }{3\phi }-\frac{H\dot \phi }\phi +\frac{\omega (\phi
)\dot \phi ^2}{6\phi ^2}-\frac k{
\begin{array}{c}
\begin{array}{c}
a^2 \\ 
\end{array}
\\ 
\end{array}
}
\end{equation}
\begin{equation}
\ddot \phi +3H\dot \phi +\frac{\omega ^{\prime }\dot \phi ^2}{2\omega +3}=%
\frac{8\pi \rho (4-3\gamma )}{2\omega +3}
\end{equation}
\begin{equation}
\rho \propto a^{-3\gamma }.
\end{equation}
We will now focus attention upon solutions of these equations with $k=0$ in
the case of Brans-Dicke theory ($\omega $ constant). Scalar-tensor theories
with varying $G$ differ from general relativity in that they admit vacuum
solutions when $k=0$ (O'Hanlon \& Tupper 1970)

\begin{eqnarray}
G &\propto &\phi ^{-1}\propto t^{-d/(1+d)} \\
a(t) &\propto &t^{1/3(1+d)}  \nonumber \\
d &\equiv &\omega ^{-1}\left( 1+\sqrt{1+\frac{2\omega }3}\right) .  \nonumber
\end{eqnarray}
They also possess a class of special power-law solutions for perfect-fluid
universes, (Nariai 1969),

\begin{eqnarray}
G &\propto &\phi ^{-1}\propto t^{-B} \\
a(t) &\propto &t^A  \nonumber \\
p &=&(\gamma -1)\rho  \nonumber \\
&&  \nonumber
\end{eqnarray}
where

\begin{equation}
A=\frac{2+2\omega (2-\gamma )}{%
\begin{array}{c}
4+3\omega \gamma (2-\gamma ) \\ 
\end{array}
},
\end{equation}
\begin{equation}
B=\frac{2(4-3\gamma )}{%
\begin{array}{c}
4+3\omega \gamma (2-\gamma ) \\ 
\end{array}
}.
\end{equation}
The general solutions can be found for all $\gamma $ but are rather
cumbersome and opaque; exact solutions for $k=0$ have been found by
Gurevich, Finkelstein and Ruban (1973) and for all $k$ by Barrow (1993a); a
phase plane analysis has been performed by Kolitch and Eardley (1995) which
includes the $k\neq 0$ models. Their general properties are as follows. As $%
t\rightarrow 0$ they approach the vacuum solutions (92), while as $%
t\rightarrow \infty $ they approach the matter-dominated solutions
(93)-(95). As $\omega \rightarrow \infty $ these matter-dominated solutions
approach the general relativistic results, $a(t)\propto t^{2/3\gamma }$, $G$
constant. There is a smooth transition between these simple early and late
time behaviours. Thus the power-law matter-dominated solutions (93)-(95) are
unstable as $t\rightarrow 0$. The general solutions are dominated by the
Brans-Dicke scalar field. A similar early-time behaviour occurs in the $%
\omega (\phi )$ theories although the form of the early vacuum-dominated
phase depends on the detailed functional form of $\omega (\phi )$, (see
Barrow 1993a, 1993b, Barrow and Mimoso 1994, Damour and Nordvedt 1993, Serna
and Alimi 1996, and Barrow and Parsons 1996 for details).

\subsection{New Methods of Solution}

\subsubsection{Vacuum and radiation models}

The general solutions to Eqs.~(52) - (54) contain four arbitrary integration
constants, one more than their GR counterparts, the extra degree of freedom
being attached to the value of $\dot \phi $. When the energy-momentum tensor
is trace-free there exists a conformal equivalence between the theory and
GR, the right-hand side of Eq.~(53) vanishes and $\dot \phi =0$ is always a
particular solution, corresponding to a special choice of the additional
constant possessed by the model over GR. Consequently, the exact general
solution of Einstein's equations when $T_{\mathrm{ab}}$ is trace-free is
also a particular solution to Eqs.~(52)--(54) with $\phi $, and hence $%
\omega (\phi )$, constant.

It will seldom be the case that the particular solution obtained in this way
will form the general solution for that particular choice of $\omega (\phi )$
. Usually, however, it will be the late or early time attractor of the
general solution. For example, in the case of Brans-Dicke theory the special
GR solution is the late-time attractor for flat and open universes but not
the early-time attractor. However, a method has been developed for
integrating the field equations for models containing trace-free matter
(Barrow 1993a). The procedure is as follows.

Eq.~(54) integrates immediately to yield 
\begin{equation}
8\pi \rho =3\Gamma a^{-3\gamma }\,,  \label{cos5}
\end{equation}
where $\Gamma \geq 0$ is a constant of integration; $\Gamma =0$ describes
vacuum models. Making the choice $\gamma =4/3$, corresponding to blackbody
radiation, and introducing the conformal time co-ordinate, $\eta $, defined
by 
\begin{equation}
ad\eta =dt\,,  \label{conf}
\end{equation}
Eq.~(53) becomes 
\begin{equation}
\phi _{\eta \eta }+\frac 2aa_\eta \phi _\eta =-\frac{\omega ^{\prime }(\phi )%
}{2\omega (\phi )+3}\left( \phi _\eta \right) ^2\,,
\end{equation}
where subscript $\eta $ denotes a derivative with respect to conformal time.
This integrates exactly to give, 
\begin{equation}
\phi _\eta a^2=3^{1/2}A(2\omega (\phi )+3)^{-1/2}\,;\;\;\;A\;\;\mathrm{const.%
}  \label{phip}
\end{equation}
We now employ the variable (suggested by the conformal invariance) used by
Lorenz-Petzold (1984) to study Brans-Dicke models, 
\begin{equation}
y=\phi a^2\,,  \label{ydef}
\end{equation}
to re-write the scalar-tensor version of the Friedmann equation, Eq.~(52),
as 
\begin{equation}
\left( y_\eta \right) ^2=-4ky^2+4\Gamma y+\frac 13\left( \phi _\eta \right)
^2a^4(2\omega (\phi )+3)\,.  \label{ycom}
\end{equation}
Dividing Eq.~(\ref{phip}) by Eq.~(\ref{ydef}), and using Eq.~(\ref{phip}),
we obtain the coupled pair of differential equations 
\begin{eqnarray}
\frac{\phi _\eta }\phi &=&3^{1/2}Ay^{-1}(2\omega (\phi )+3)^{-1/2}\,,
\label{trace2} \\
\left( y_\eta \right) ^2 &=&-4ky^2+4\Gamma y+A^2\,.
\end{eqnarray}
We may now obtain the general solution for a particular choice of $\omega
(\phi )$, given $k$. Integrating Eq.~(\ref{trace2}) yields $y(\eta )$ which,
in conjunction with $\omega (\phi )$, implies $\phi (\eta )$ and, without
further integration, $a(\eta )$ from Eq.~(\ref{ydef}). If Eq.~(\ref{conf})
can be integrated and inverted we may compute $\phi (t)$ and $a(t)$, so
completing the solution. The vacuum models are obtained by setting $\Gamma
=0.$

\subsubsection{General perfect-fluid cosmologies}

When $T,$the trace of the energy-momentum tensor is non-vanishing, the
situation is substantially more complicated. In this instance, $\dot \phi =0$
is no longer a particular solution of the field equations, forcing us to
resort to more elaborate methods to obtain solutions. Barrow and
Mimoso(1994) have done this, for the $k=0$ models, by generalising the
method of Gurevich \emph{et al.} (1973) for BD models to the case of
non-constant $\omega (\phi )$. We outline this procedure.

Introducing the new time co-ordinate $\xi $, and the two new variables $x$
and $v$ such that 
\begin{equation}  \label{timeperf}
dt=a^{3(\gamma -1)}\;\sqrt{\frac{2\omega +3}3}\;d\xi \quad ,
\label{edeftimevar}
\end{equation}
\begin{equation}
x\equiv \left[ \phi a^{3(1-\gamma )}\left( a^3\right) _\xi \right] \;,
\label{edefx}
\end{equation}
\begin{equation}
v\equiv \left[ a^{3(2-\gamma )}\phi _\xi \right] \quad ,  \label{edefy}
\end{equation}
and confining attention to the $k=0$ models, Eqs.~(52)-(54) transform to 
\begin{equation}
\left( \frac 23x+v\right) ^2=\left( \frac{2\omega +3}3\right) \,\left[
v^2+4\Gamma \,\phi \,a^{3(2-\gamma )}\right] \;,  \label{eFriedm}
\end{equation}
\begin{equation}
v_\xi =\Gamma \;(4-3\gamma )\;,  \label{edy}
\end{equation}
and 
\begin{equation}
x_\xi =3\Gamma \,\left[ (2-\gamma )\omega +1\right] +\frac 32\left( \frac
2{3(2\omega +3)}x+v\right) \,\omega _\xi \;,  \label{ex'}
\end{equation}
where subscript-$\xi $ represents a derivative with respect to $\xi $-time.
Eqs.~(\ref{edy}) and (\ref{ex'}) integrate easily to yield 
\begin{eqnarray}
v &=&\Gamma (4-3\gamma )\,(\xi -\xi _1)\quad ,  \label{eyeta} \\
x &=&\frac 32\;\left[ -v+\sqrt{2\omega +3}\left( C+\Gamma (2-\gamma
)\,\int_{\xi _1}^\xi \,\sqrt{2\omega +3}\,d\bar \xi \right) \right] \quad ,
\label{exeta}
\end{eqnarray}
$C$ is an integration constant and $\xi _1$ fixes the origin of $\xi $-time.
Noting the relation 
\begin{equation}
\frac 3{a\phi }a_\xi \phi _\xi =\frac 1{\phi ^2}\left( \phi _\xi \right)
^2\;3\frac \phi a\frac{a_\xi }{\phi _\xi }=\frac 1{\phi ^2}\,\left( \phi
_\xi \right) ^2\frac xv\;,  \label{erelxy}
\end{equation}
and differentiating $y$, with respect to $\xi $, yields 
\begin{equation}
\left( \frac{\phi _\xi }\phi \right) _\xi +\left[ \frac{3\gamma -4}2+\frac
1{\xi -\xi _1}f_\xi (\xi )\right] \;\left( \frac{\phi _\xi }\phi \right)
^2=\frac 1{\xi -\xi _1}\,\frac{\phi _\xi }\phi \quad ,  \label{ephi''}
\end{equation}
where a new function $f(\xi )$, is defined by 
\begin{equation}
f(\xi )\equiv \int_{\xi _1}^\xi \,\frac{3(2-\gamma )}{2\Gamma (4-3\gamma )}\,%
\sqrt{2\omega (\phi )+3}\;\left[ C+\Gamma (2-\gamma )\,\int_{\xi _1}^\xi \,%
\sqrt{2\omega (\phi )+3}\,d{\tilde \xi }\right] \,d\bar \xi \;.
\label{edeff}
\end{equation}
Solving Eq.~(\ref{ephi''}), we have the solution 
\begin{equation}  \label{gdef}
\ln {\left( \frac \phi {\phi _0}\right) }=\int_{\xi _1}^\xi \,\frac{\xi -\xi
_1}{g(\xi )}d\xi \,,  \label{elnphi}
\end{equation}
with $g(\xi )$ simply related to $f(\xi )$ by 
\begin{equation}
g(\xi )\equiv f(\xi )+\frac{3\gamma -4}4\,(\xi -\xi _1)^2+D\,,\quad
\label{edefg}
\end{equation}
where $D$ is a constant of integration. Eq.~(\ref{erelxy}) immediately
reveals a simple formula for the scale-factor: 
\begin{equation}  \label{aofg}
a^3=a_0^3\;\left( \frac g\phi \right) ^{\frac 1{2-\gamma }}\;;a_0\;\mathrm{%
constant}\,.  \label{eV}
\end{equation}
Finally, the scalar-tensor coupling function $\omega (\phi )$ is given as a
function of $f$ by 
\begin{equation}
2\omega \left( \phi (\xi )\right) +3=\frac{4-3\gamma }{3(2-\gamma )^2}\;%
\frac{(f^{\prime })^2}{\left[ f+\frac{4-3\gamma }{3(2-\gamma )^2}%
\,f_0\right] }\quad ,  \label{edefomega}
\end{equation}
where $f_0$ is another arbitrary constant.

An important benchmark is provided by the behaviour of the BD theory, where $%
\omega (\phi )=\omega _0=$ constant. In this case, the generating function, $%
f(\xi )$, is given by a quadratic in $\xi $: 
\begin{equation}
f_{\mathrm{BD}}(\xi )=\frac{3(2-\gamma )}{2\Gamma (4-3\gamma )}\,\sqrt{%
2\omega _0+3}\;\left[ C\;\xi +\frac{\Gamma (2-\gamma )}2\,\xi ^2\;\sqrt{%
2\omega _0+3}\right] \,.  \label{eg(eta)JBD}
\end{equation}
Hence, in general $(C\neq 0\neq \Gamma )$ when $\gamma \neq 4/3\,,\;2$, we
see that $f_{\mathrm{BD}}\propto \xi ^2$ as $\xi \rightarrow \infty $ and $%
f_{\mathrm{BD}}\propto \xi $ as $\xi \rightarrow 0$, where $dt\propto
a^{3(\gamma -1)}d\xi $. If we choose $C=0$ then $f_{\mathrm{BD}}\propto
\Gamma \xi ^2$ as $\xi \rightarrow 0$. The choice $C=0$ restricts the
solution to the special `matter-dominated' solutions (termed `Machian' by
Dicke, see Weinberg (1972)) which were first found for all perfect-fluids by
Nariai (1969). If $C\neq 0$ then the early-time behaviour is dominated by
the dynamics of the $\phi $-field; such solutions are termed `$\phi $%
-dominated' (or `non-Machian' by Dicke).

Therefore if we choose a generating function $g(\xi )$ that grows slower
than $\xi ^2$ as $\xi \rightarrow \infty $ it will produce a theory that
approaches BD at late times ($\phi \rightarrow $constant, $\omega (\phi
)\rightarrow $ constant), whilst if $g(\xi )$ decreases slower than $\xi $
as $\xi \rightarrow 0$ then the theory will approach the behaviour of $\phi $%
-dominated BD theory at early-times. This means that we will find new
(non-BD) late-time behaviours by studying generating functions which
increase faster than $g(\xi )=\xi ^2$ as $\xi \rightarrow \infty $ and new
(non-BD) early-time behaviour by picking generating functions which decrease
slower than $g(\xi )=\xi $ as $\xi \rightarrow 0$ or $\xi \rightarrow \xi _{%
\mathrm{min}}$ (if there is no zero of $\xi $ at the minimum of $a(t)$).

\subsection{Comparing Newtonian and Relativistic Cosmologies}

It is now possible to consider some of the similarities and differences that
exist between the Newtonian cosmologies with varying $G$ and their curved
space-time counterparts. We recall that the special Newtonian power-law
solutions (24) have the form $G\propto t^{-n}$ with scale factor $r\propto
t^{(2-n)/3\gamma }.$ This is identical to the Brans-Dicke solution when $p=0$
and

\begin{equation}
n=\frac 2{4+3\omega }.
\end{equation}
In fact, for any choice of $\omega (\phi )$ dust universes have

\begin{equation}
\phi a^3\rightarrow t^2
\end{equation}
which corresponds to Newtonian solutions with $G\propto t^{-n}$ and scale
factor $r\propto t^{(2-n)/3\text{ }\ }$.

The vacuum solutions have a slightly different structure. If we identify

\begin{equation}
n=\frac d{d+1}
\end{equation}
then we have

\begin{equation}
a(t)\propto t^{1/3(1+d)}\propto t^{(1-n)/3}.
\end{equation}

However, there is no general correspondence for other equations of state.
Most notably, Brans-Dicke radiation-dominated universes ($\gamma =4/3)$ are
the same as in general relativity, $G=$ constant and $a(t)\propto t^{1/2},\ $%
and differ from the radiation solutions. In general, the Newtonian solutions
are the same as the Brans-Dicke matter-dominated solutions only if we make
the choice

\begin{equation}
n=\frac{2(4-3\gamma )}{4+3\omega \gamma (2-\gamma )}.
\end{equation}
Also, as can be seen from the right-hand side of the $\phi (t)$ evolution
equation, (90), the effective sign of the gravitational coupling changes
sign from negative to positive when $\gamma $ becomes greater than 4/3.

The Newtonian inflationary solutions with $\gamma =0$, given in eqn. (29),
do not have direct counterparts in Brans-Dicke gravity theories. However,
scalar tensor theories with

\begin{equation}
2\omega +3\propto \phi ^h
\end{equation}
have solutions of similar form as $t\rightarrow \infty $ with (Barrow and
Mimoso 1994, Barrow and Parsons 1996)

\begin{equation}
G\propto \phi ^{-1}\propto t^{-2/(2h+1)}
\end{equation}
and the scale factor evolves as

\begin{equation}
a(t)\propto t^{(h-1)/3(2h+1)}\exp \left\{ At^{2h/(2h+1)}\right\} .
\end{equation}

Thus, with $2h+1>0,$ we have $\omega \rightarrow \infty $ and $\omega
^{\prime }/\omega ^3\rightarrow 0$ as $t\rightarrow \infty $ and general
relativity is approached in the weak-field limit. These solutions are not of
identical form to the Newtonian solutions with $G\propto t^{-n}$ and $%
n=2/(2h+1).$

\section{Gravitational memory}

The process of primordial black hole formation in a cosmological model with
varying $G$ creates an interesting problem. We know (Hawking 1972) that
black holes in scalar-tensor gravity theories are identical to those
occurring in general relativity. Suppose, for simplicity, that a
Schwarzschild black hole forms in the very early universe at a time $t_f$
when the gravitational coupling $G(t_f)$ differs from the value $G(t_0)$
that we observe at the present cosmic time, $t_0.$ This black hole will have
an horizon size $R_f=2G(t_f)M\sim t_f$ when it forms. We now ask what
happens to this black hole during the subsequent evolution of the universe
as the value of $G$ changes with time in the background universe (Barrow
1992, Barrow and Carr 1996).

There appear to be two alternatives. The value of $G(t)$ on the scale of the
even horizon could change at the same rate as that in the background
universe. But this would imply that the black hole was changing with time.
Thus it could not be one of the black holes defined in general relativity,
as required. Moreover, if $\dot G<0,$the horizon area ($A_{hor}$) would
decrease with time and so would the associated entropy which is given by
(Kang 1996)

\[
S_{bh}=\frac{A_{hor}\times \phi _{hor}}4 
\]
where $\phi _{hor}=G_{hor}^{-1}$ determines the value of $G$ on the horizon.
Alternatively, there might be a process of 'gravitational memory' (Barrow
1992) wherein the scalar field determining $G$ remains constant on the scale
of the black hole horizon whilst changing in the cosmological background.
That is at any moment of cosmic time there would be a space variation in $G.$
This has dramatic implications for the Hawking evaporation of primordial
black holes. For the lifetime and temperature of an evaporating black hole
will be determined by the value of $G(t_f)$ at the time when it formed
rather than by the value $G(t_0)$ we observe in the universe today: the
black hole 'remembers' the value of $G$ at the time of its formation. Hence,
its Hawking lifetime will be $\tau _{bh}\sim G_f^2M^3$ and its temperature $%
T_{bh}\sim G_f^{-1}M^{-1}.$

Black holes which explode today are those whose Hawking lifetime is equal to
the present age of the universe. This fixes their masses to be

\[
M_{ex}\simeq 4\times 10^{14}\times \left( \frac{G(t_0)}{G(t_f)}\right)
^{\frac 23}gm 
\]
and their temperature when the explode is therefore given by

\[
T_{ex}\simeq 24\times \left( \frac{G(t_0)}{G(t_f)}\right) ^{\frac 13}MeV. 
\]

Clearly, if there has been significant change in the value of $G$ since $%
t_f\sim 10^{-23}s$ then the physical characteristics of exploding black
holes can be very different from those predicted under the assumption that $%
G $ does not change (Hawking 1974). Quite modest amounts of time evolution
at unobservably early times can shift the spectral range in which we would
see the evaporation products out of the gamma-ray band. This means that we
should be looking for the evidence of black hole evaporation in other parts
of the electromagnetic spectrum. A more detailed study of the observational
evidence, seen is this light, is given by Barrow and Carr (1996). 

\section{Origins of the Values of Constants}

In recent years there has been a good deal of speculation about mechanisms
for explaining the values of the fundamental constants. At one time it was
widely believed that some ultimate Theory of Everything would eventually
tell us that the constants could have one, and only one, set of logically
self-consistent values. Such a simple scenario now seems less and less
likely (Barrow 1991). There are so many sources of randomness in the process
which endow the fundamental constants with their low-energy values, and the
parameters likely to be fixed by a Theory of Everything are so far removed
from our three-dimensional physical constants, that many new possibilities
must be taken seriously. The non-uniqueness of the ground state of any
Theory of Everything would mean that fundamental constants could take on
many self-consistent sets of values. We would have to use anthropic
constraints in order to understand those that we observe. This creates new
interpretational problems. Our underlying Theory of Everything would have
quantum gravitational characteristics and its predictions about constants
would have a probabilistic form. Although, formally, there would be a most
probable value for the low-energy measurement of a quantity like the fine
structure constant, such a value might be irrelevant for observational
purposes (Barrow 1994). We would only be interested in the range of values
for which the evolution of complexity, in the form that we call 'life', is
possible. This may well confine us to a subset of values which, a priori, is
extremely improbable. This shows that in order to make a correct comparison
of the probabilistic predictions of such a theory with observation we would
need to know every dependence of processes which can lead to the evolution
of complexity on the values of the constants of Nature.

The fact that the observed values of many of the constants of Nature fall
within a very narrow range for which life appears to be possible has
elicited a variety of interpretations:

(i) \textbf{Good luck:} the constants are what they are and could be no
other way. The range which allows intelligent observers to evolve and
persist is narrow and we are very lucky that our universe falls within that
range. No matter how improbable this sate of affairs we could observe it to
be no other way.

(ii) \textbf{Life is inevitable:} we have been misled by our limited
knowledge of complexity into thinking that life is restricted to universes
spanned by a very narrow range of values for the constants. In fact, life
may be a widespread inevitability in the phase space of all possible values
of the constants. Even our own form of carbon-based life may exist in other
novel forms which exploit the possibilities provided by the recently
discovered fullerene chemistry. complexity of the sort that we call life may
also exist in quite different forms to those we are accustomed to: for
example, existing in velocity space rather than in position space.

(iii) \textbf{All possibilities exist:} whether through the actualisation of
quantum-mechanical many-worlds, the realisation of all logically consistent
Theories of Everything, or some elaboration of the self-reproducing universe
scenarios, every possible permutation of the values of the constants exists
in some universe. We live in one of the subset which allows life to exist.
It is also possible that the ensemble of possibilities is played out in a
single infinite Universe and we inhabit one of the life-supporting parts of
it.

(iv)\textbf{\ Cosmic fine tuning:} some physical process brings about
approach to a particular set of values for the constants over long periods
of time, perhaps through many cycles of cosmic evolution. The attracting set
may be predictable in certain respects.

The last of these four possibilities has attracted some interest recently.
Harrison (1995) made the amusing suggestion that the fine tuning of the
constants may be the end result of successive intelligent interventions by
beings able to create universes in the laboratory (something discussed in
the literature even by ourselves! Farhi and Guth (1987)).Aware that certain
combinations of the values of fundamental constants raise the probability of
life evolving, and able to engineer these values at inception, successive
generations would tend to find themselves inhabiting universes in which
life-supporting combinations obtained to high precision. Although Harrison
refers to his as a 'natural selection' of universes, it is more akin to
artificial selection, or forced breeding.

Linde (1990) has proposed generalisations of the self-reproducing eternal
inflationary universe in which the values of the fundamental constants
change from generation to generation. Although unobservable, this scheme has
the merit of being a by-product of the standard chaotic inflationary
universe scenario.

A third scenario of this sort, which has attracted a surprising amount of
attention is that proposed by Smolin (1992) who suggested that a bounce, or
quantum tunnelling, occurs at all final black-hole-collapse singularities
which transforms them into initial singularities for new expanding
universes. During this process the constants of Nature undergo small random
changes. It is expected therefore that selection pressure will act so as to
maximise the black holes produced in universes as time goes on (no weighting
of the volume taking part in this reproduction process is introduced though,
as is the case in the self-reproducing inflationary universes). Thus, if our
universe is the result of the action of this selection process over many
cycles of collapse and re-expansion, in which the constants have lost memory
of any initial conditions they may have had, then Smolin argues that we
would expect to be near a local maximum in the black-hole production. Hence, 
\textit{small} changes in the constants of Nature should in general take us
downhill from this local maximum and always \textit{reduce} the amount of
black hole production. By conducting such thought-experiments the general
consistency of the idea can be tested.

We make three remarks about this speculative scenario. First, it is not
clear that is as sharply predictive and testable as claimed. We should only
expect to find ourselves residing near a local maximum in the space of
constants if that maximum also provides conditions which permit living
observers to exist. If those condition are unusual then we might have to
exist in one of the improbable universes far from the local maxima. We can
only determine if this is the case by having a complete understanding of the
necessary and sufficient conditions for living complexity to exist. Second,
putting this objection to one side, there may well be small changes in the
values of the constants which significantly increase the production of black
holes. For example a small ($70KeV$) strengthening of the strong interaction
would bind the dineutron and the diproton (helium-2), so providing a direct $%
H+H\rightarrow He^2$ channel for nuclear burning. massive stars would run
through their evolution very rapidly and end as black holes far sooner and
with higher likelihood than at present (see Dyson 1971, Barrow 1987). Third,
we might ask why there should be any local maxima at all for variations in
certain constants. Variation would proceed to states of higher gravitational
entropy by always increasing the value of $S_{bh}\propto GM^2,$ and this
would be effected by a random walk upwards through (over long time averaged)
increasing values of $G.$

Finally, we should note that any scheme which relies upon random changes in
the constants of Nature occurring at the endpoint of gravitational collapse
must beware of the consequences of changes which prevent future collapses
from occurring. A specific example is seen in the case of closed universes
oscillating under the requirement that their total entropy increase from
cycle to cycle. There, one finds that any positive cosmological constant (no
matter how small in value) which remains constant (or falls slowly enough on
average) from cycle to cycle ultimately stops the sequence of growing
oscillations and leaves the Universe in a state of indefinite expansion
which asymptotes towards the de Sitter state (Barrow and Dabrowski 1995). In
Smolin's scenario one might consider that if the curvature of space or the
value of the cosmological constant, or the magnitude of vacuum stresses
associated with scalar fields which violate the strong energy condition,
were to change at the collapse event in ways that prevented future collapse
of some or all of the Universe, then gradually the fraction of the Universe
which could gravitationally collapse and evolve the values of its constants
by random reprocessing would shrink asymptotically to zero. Evolution would
cease. This Universe would have 'died'.

\section{Simultaneous Variations of Many Constants}

The subject of varying constants is of particular current interest because
of the new possibilities opened up by the structure of unified theories,
like string theory and M-theory, which lead us to expect that additional
compact dimensions of space may exist. Although these theories do not
require traditional constants to vary, they allow a rigorous description of
any variations to be provided: one which does not merely `write in' the
variation of constants into formulae derived under the assumption that they
do not vary. This self-consistency is possible because of the presence of
extra dimensions of space in these theories. The `constants' seen in a
three-dimensional subspace of the theory will vary at the same rate as any
change occurring in the extra compact dimensions. In this way, consistent
simultaneous variations of different constants can be described and searches
for varying constants provide a possible observational handle on the
question of whether extra dimensions exist (Marciano, 1984, Barrow 1987,
Damour \& Polyakov 1994).

Prior to the advent of theories of this sort, only the time variation of the
gravitational constant could be consistently described using scalar-tensor
gravity theories, of which the Brans-Dicke theory is the simplest example.
The modelling of variations in other `constants' was invariably carried out
by assuming that the time variation of a constant quantity, like the fine
structure constant, could just be written into the usual formulae that hold
when it is constant. One way of avoiding this situation is to exploit the
invariance properties of the non-relativistic Schr\"odinger equation for
atomic structure, which allow it to be written in dimensionless form when
atomic ('Hartree') units are chosen. It can be shown (Barrow and Tipler
1986) that any solution with an energy eigenvalue $E$, arising when the fine
structure constant is $\alpha $ and the electron mass is $m_e,$must be
related to a solution defined by a $E,\alpha ^{\prime },$ and $m_e^{^{\prime
}}$ by the relation

\begin{equation}
\frac E{\alpha ^2m_ec^2}=\frac{E^{\prime }}{\alpha ^{\prime 2}m_e^{\prime
}c^2}  \label{eq.1}
\end{equation}
where $c$ is the velocity of light.

The possibility of linked variations in low-energy constants as a result of
high-energy unification schemes has the added attraction of providing a more
powerful means of testing those theories (Marciano 1984, Kolb, Perry, \&
Walker, 1986, Barrow 1987, Dixit \& Sher 1988, Campbell \& Olive 1995).

Higher-dimensional theories typically give rise to relationships of the
following sort 
\begin{eqnarray}
\alpha _i(m_{*}) &=&A_iGm_{*}^2=B_i\lambda ^n(\ell _{pl}/R)^k;n,k\text{ }%
\mathrm{constants}  \label{ } \\
\alpha _i^{-1}(\mu ) &=&\alpha _i^{-1}(m_{*})-  \nonumber \\
\lefteqn{\pi ^{-1}\sum C_{ij}[\ln (m_{*}/m_j)+\theta (\mu -m_j)\ln (m_j/\mu
)]+\Delta _i}  \nonumber
\end{eqnarray}
where $\alpha _i(..)$ are the three gauge couplings evaluated at the
corresponding mass scale; $\mu $ is an arbitrary reference mass scale, $%
m_{*} $ is a characteristic mass scale defining the theory (for example, the
string scale in a heterotic string theory); $\lambda $ is some dimensionless
string coupling; $\ell _{pl}=G^{-1/2}$ is the Planck length, and $R$ is a
characteristic mean radius of the compact extra-dimensional manifold; $%
C_{ij} $ are numbers defined by the particular theories and the constants $%
A_i$ and $B_i$ depend upon the topology of the additional ( $>3$)
dimensions. The sum is over $j=$ leptons, quarks, gluons, $W^{\pm },\ Z$ and
applies at energies above $\mu \sim 1GeV$ (Marciano 1984)$.$ The term $%
\Delta _i$ corresponds to some collection of string threshold corrections
that arise in particular string theories or an over-arching M theory
(Antoniadis \& Quiros 1996). They contain geometrical and topological
factors which are specified by the choice of theory. By differentiating
these two expressions with respect to time (or space), it is possible to
determine the range of self-consistent variations that are allowed. In
general, for a wide range of super-symmetric unified theories, the time
variation of different low-energy constants will be linked by a relationship
of the form (where we consider $\dot \beta $ to denote the time derivative
of $\beta $ etc.) 
\begin{equation}
\delta _0\frac{\dot \beta }\beta =\delta _1\frac{\dot G}G+\sum \delta _{2i}%
\frac{\dot \alpha _i}{\alpha _i^2}+\delta _3\frac{\dot m_{*}}{m_{*}}+\sum
\delta _{4j}\frac{\dot m_j}{m_j}+\delta _5\frac{\dot \lambda }\lambda +...\ 
\end{equation}
where $\beta \equiv m_e/m_{pr}$ (Drinkwater et al 1997). It is natural to
expect that all the terms involving time derivatives of `constants' will
appear in this relation unless the constant $\delta $ prefactors vanish
because of supersymmetry or some other special symmetry of the underlying
theory. This relation shows that, since we might expect all terms to be of
similar order (although there may be vanishing constant $\delta $ prefactors
in particular theories), we might expect variations in the Newtonian
gravitational `constant', $\dot G/G$, to be of order $\dot \alpha /\alpha
^2. $

\section{Varying alpha -- New Observational Limits}

Quasar absorption systems present ideal laboratories in which to search for
any temporal or spatial variation in the assumed fundamental constants of
Nature. Such ideas date back to the 1930s, with the first constraints from
spectroscopy of QSO absorption systems arising in the 1960s. An historical
summary of the various propositions is given in Varshalovich \& Potekhin
(1995) and further discussion of their theoretical consequences is given in
Barrow \& Tipler (1986).

Recently, we have considered the bounds that can be placed on the variation
of the fine structure constant and proton $g\ $factor from radio
observations of atomic and molecular transitions in high redshift quasars
(Drinkwater et al 1997). To do this we exploited the recent dramatic
increase in quality of spectroscopic molecular absorption at radio
frequencies, of gas clouds at intermediate redshift, seen against background
radio-loud quasars. Elsewhere, we will consider the implications of
simultaneous variations of several `constants' and show how these
observational limits can be used to constrain a class of inflationary
universe theories in which small fluctuations in the fine-structure constant
are predicted to occur.

The rotational transition frequencies of diatomic molecules such as CO are
proportional to ${\hbar /( Ma^2)}$ where $M$ is the reduced mass and $a={%
\hbar^2 /( m_e e^2)}$ is the Bohr radius. The 21\textrm{\thinspace cm}\
hyperfine transition in hydrogen has a frequency proportional to ${\mu_p
\mu_B /( \hbar a^3)}$, where $\mu_p = g_p {e \hbar / (4m_p c)}$, $g_p$ is
the proton g-factor and $\mu_B = {e \hbar / (2m_e c)}$. Consequently
(assuming $m_p/M$ is constant) the ratio of a hyperfine frequency to a
molecular rotational frequency is proportional to $g_p \alpha^2$ where $%
\alpha = {e^2 / ( \hbar c)}$ is the fine structure constant. Any variation
in $y\equiv g_p \alpha^2$ would therefore be observed as a difference in the
apparent redshifts: ${\Delta z / (1+z)} \approx {\Delta y / y}$. Redshifted
molecular emission is hard to detect but absorption can be detected to quite
high redshifts (see review by Combes \& Wiklind, 1996). Recent measurements
of molecular absorption in some radio sources corresponding to known HI 21%
\textrm{\thinspace cm}\ absorption systems give us the necessary combination
to measure this ratio at different epochs.

Common molecular and HI 21\textrm{\thinspace cm}\ absorptions in the radio
source PKS 1413+135 have previously been studied by Varshalovich \& Potekhin
(1996). They reported a difference in the redshifts of the CO molecular and
HI 21\textrm{\thinspace cm}\ atomic absorptions which they interpreted as a
mass change of $\Delta M/M=(-4\pm 6)\times 10^{-5}$ but as we show above
this comparison actually constrains $g_p\alpha ^2$, not mass. Furthermore
they used overestimates of both the value and error. They used the Wiklind
\& Combes (1994) measurement which had the CO line offset from the HI
velocity by $-$11 kms$^{-1}$; a corrected CO measurement (Combes \& Wiklind,
1996) shows there is no measurable offset. Furthermore, Varshalovich \&
Potekhin (1996) used the width of the HI line for the measurement
uncertainty. Even allowing for systematic errors the true uncertainty is at
least a factor of 10 smaller so these data in fact establish a limit of
order $10^{-5}$ or better. This potential for improved limits has prompted
the present investigation: previous upper limits on change in $\alpha $ are
of order $\Delta \alpha /\alpha \approx 10^{-4}$ (Cowie \& Songalia, 1995;
Varshalovich, Panchuk \& Ivanchik, 1996).

\subsection{Comparison of HI and molecular systems}

Our new more accurate redshift estimates (Drinkwater et al 1997) give a
molecular redshift of $0.684680\pm 0.000006$ and an a 21cm redshift of $%
0.684684\pm 0.000006$ for the source $0218+357$ and a molecular redshift of $%
0.246710\pm 0.000005$ and a 21cm redshift of $0.246710\pm 0.000004$ for the
source 1413+135. We can therefore combine the uncertainties in quadrature to
give 1-sigma upper limits on the redshift differences. These give $|{\Delta
z/1+z}|<5\times 10^{-6}$ (1.5 $kms^{-1}$) for both sources.

We must still consider the possibility that the molecular and atomic
absorption arises in different gas clouds along the line of sight. This
could explain any observed difference. However there is no measurable
difference between the two velocities in our data, so we are probably
detecting the same gas. The alternative would be that there was a change in
the frequencies but that in both cases it was exactly balanced by the random
relative velocity of the two gas clouds observed. We consider this very
unlikely because of the small 1Kms$^{-1}$\ dispersion within single clouds.~

We can now use the limits to ${\Delta z/1+z}$ with the relationship to
derive 1-sigma limits on any change in $y=g_p\alpha ^2$: $|{\Delta y/y}%
|<5\times 10^{-6}$ at both $z=0.25$ and $z=0.68$. These are significantly
lower than the previous best limit of $1\times 10^{-4}$ by Varshalovich \&
Potekhin (1996) (it was quoted as a limit on nucleon mass, but it actually
refers to $g_p\alpha ^2$).

As there are no theoretical grounds to expect that the changes in $g_p$ and $%
\alpha ^2$ are inversely proportional, we obtain independent rate-of-change
limits of $|\dot {g_p}/g_p|<2\times 10^{-15}\mathrm{\,y}^{-1}$ and $|\dot
\alpha /\alpha |<1\times 10^{-15}\mathrm{\,y}^{-1}$ at $z=0.25$ and $|\dot
{g_p}/g_p|<1\times 10^{-15}\mathrm{\,y}^{-1}$ and $|\dot \alpha /\alpha
|<5\times 10^{-16}\mathrm{\,y}^{-1}$ at $z=0.68$ (for $H_0=75$\thinspace
km\thinspace s$^{-1}$\thinspace Mpc$^{-1}$ and $q_0=0$ assumed here). These
new limits are stronger than the previous 1 sigma limit of $|\dot \alpha
/\alpha |<8\times 10^{-15}\mathrm{\,y}^{-1}$ at $z\approx 3$ (Varshalovich
et al.\ 1996).

The most stringent laboratory bound on the time variation of $\alpha $ comes
from a comparison of hyperfine transitions in Hydrogen and Mercury atoms
(Prestage et al.\ 1995), $\left| \dot \alpha /\alpha \right| <3.7\times
10^{-14}y^{-1},$ and is significantly weaker than our astronomical limit.
The other strong terrestrial limit that we have on time variation in $\alpha 
$ comes from the analysis of the Oklo natural reactor at the present site of
an open-pit Uranium mine in Gabon, West Africa. A distinctive thermal
neutron capture resonance must have been in place 1.8 billion years ago when
a combination of fortuitous geological conditions enriched the subterranean
Uranium-235 and water concentrations to levels that enabled spontaneous
nuclear chain reactions to occur (Maurette 1972). Shlyakhter (1976, 1983)
used this evidence to conclude that the neutron resonance could not have
shifted from its present specification by more than $5\times 10^{-4}eV$ over
the last 1.8 billion years and, assuming a simple model for the dependence
of this energy level on coupling constants like $\alpha ,$ derived a limit
in the range of $\left| \dot \alpha /\alpha \right| <(0.5$--$1.0)\times
10^{-17}y^{-1}.$ The chain of reasoning leading to this very strong bound is
long, and involves many assumptions about the local conditions at the time
when the natural reactor ran, together with modelling of the effects of any
variations in electromagnetic, weak, and strong couplings. Recently, Damour
\& Dyson (1997) have provided a detailed reanalysis in order to place this
limit on a more secure foundation. They weaken Shlyakhter's limits slightly
but give a 95\% confidence limit of $-6.7\times 10^{-17}y^{-1}<\dot \alpha
/\alpha <5.0\times 10^{-17}y^{-1}.$ However, if there exist simultaneous
variations in the electron-proton mass ratio this limit can be weakened.

These limits provide stronger limits on the time variation of $\alpha $ than
the astronomical limits; however, the astronomical limits have the distinct
advantage of resting upon a very short chain of theoretical deduction and
are more closely linked to repeatable precision measurements of a simple
environment. The Oklo environment is sufficiently complex for significant
uncertainties to remain.

Unlike the Oklo limits, the astronomical limits also allow us to derive
upper limits on any \emph{spatial} variation in $\alpha .$ Spatial variation
is expected from the theoretical result that the values of the constants
would depend on local conditions and that they would therefore vary in both
time and space (Damour \& Polyakov 1994). The two sources for which we
derived limits, 0218$+$357 and 1413$+$135, are separated by 131 degrees on
the sky, so together with the terrestrial result, we find the same values of 
$\alpha $ to within $|{\Delta \alpha /\alpha }|<3\times 10^{-6}$ in three
distinct regions of the universe separated by comoving separations up to 3000%
\textrm{\thinspace Mpc}. Limits on spatial variation of $g_p\alpha ^2m_e/m_p$
were previously discussed by Pagel (1977, 1983) and Tubbs \& Wolfe (1980).
We have improved on their limits by some 2 orders of magnitude but as our
sources are at lower redshift, they are not causally disjoint from each
other.

The high-redshift measurements are now approaching the best terrestrial
measurements based on the Oklo data. These could be further improved by a
factor of 2--5 with additional observations that would not be difficult to
perform such as fitting the atomic and molecular data simultaneously,
remeasuring the HI absorptions at higher spectral resolution. 

\subsection{Inflation}

Inflation is something of a two-edged sword when it comes to discussing
variations in constants. On the one hand there are potentials with multiple
vacuum states which allow different parts of the universe to find themselves
inheriting different suites of fundamental constants, with quite different
values. On the other hand, if, as inflation leads us to expect, the whole of
our observable universe is contained within the inflated image of a single
causally connected region, then we should expect fundamental constants to
reflect that single origin and to display spatial uniformity to very high
precision. The key question is what precisely is that precision? A bench
mark for the amplitude of possible variations is provided by the amplitude
of temperature fluctuations in the microwave background, $\Delta T/T\simeq
10^{-5}$. We would expect fluctuations in the fine structure constant
created at the end of inflation to have an almost constant curvature
spectrum (because of the time-translation invariance of almost de Sitter
inflation) with an amplitude below that of $10^{-5}$. An interesting feature
of the new astronomical observations described above is that, for the first
time, they take the observational limits on spatial variations in $\alpha $
( $|{\Delta \alpha /\alpha }|<3\times 10^{-6}$) into that regime where they
may be constraining the underlying theories more strongly than are the COBE
observations. 

\textbf{Acknowledgements}

The author acknowledges support by a PPARC Senior Fellowship. Some of the
work described here was carried out in collaboration with Bernard Carr,
Michael Drinkwater, Victor Flambaum, Jos\'e Mimoso, Paul Parsons and John
Webb. I am most grateful to them for their contributions. I would also like
to thank Norma Sanchez for her encouragement and her administrative
assistants in Erice for their assistance in Erice.

\textbf{References}

Antoniadis I., Quiros M., 1996, Large Radii and String Unification, preprint.

Barker B.M., 1978, Ap. J., 219, 5

Barrow J.D., 1987, Phys. Rev., D\ 35, 1805

Barrow J.D., 1990a, in \textit{Modern Cosmology in Retrospect}, eds.,
Bertotti B., Balbinot R., Bergia S., Messina A., pp67-93, Cambridge UP,
Cambridge

Barrow J.D., 1990b, Phys. Lett., B235, 40

Barrow J.D., 1991, \textit{Theories of Everything, }Oxford UP, Oxford.

Barrow J.D., 1992, Phys. Rev. D 46, R3227

Barrow J.D., 1993a, Phys. Rev., D 47, 5329

Barrow J.D., 1993b, Phys. Rev., D 48, 3592

Barrow J.D., 1994, \textit{The Origin of the Universe, }Basic Books, NY.

Barrow J.D., 1995, Phys. Rev., D 51, 2729

Barrow J.D., 1996, MNRAS 282, 1397

Barrow J.D., Carr B.J., 1996, Phys. Rev. D54, 3920.

Barrow J.D., Dabrowski M.P.,1995 MNRAS 275, 850

Barrow J.D., G\"otz G., 1989a, Class. Quantum Gravity, 6, 1253

Barrow J.D., Liddle A.R., 1993, Phys. Rev., D 47, R5219

Barrow J.D., Maeda K., 1990, Nucl. Phys. B341, 294

Barrow J.D., Mimoso J.P., 1994, Phys. Rev., D 50, 3746

Barrow J.D., Parsons P., 1996, Phys. Rev., D 000,000

Barrow J.D., Saich P., 1990, Phys. Lett. B 249, 406

Barrow J.D., Tipler F.J., 1986, \textit{The Anthropic Cosmological Principle}%
, Oxford UP, Oxford

Batyrev A.A., 1941, Astron. Zh. 18, 343

Batyrev A.A., 1949, Astron. Zh. 26, 56

Bekenstein J.D., Meisels A., 1980, Phys. Rev., D 22, 1313

Bekenstein J.D., Milgrom M., 1984, Ap. J., 286, 7

Bekenstein J.D., Sanders R.J., 1994, Ap. J., 429, 480

Bergmann P.G., 1968, Int. J. Theo. Phys., 1, 25

Brans C., Dicke R.H., 1961, Phys. Rev.,\ 24, 925

Callan C.G., Friedan D., Martinec E.J., Perry, M.J., 1985, Nucl. Phys.,
B262, 597

Campbell B.A., Olive, K.A., 1995, Phys.Lett. B, 345, 429

Cesari L., 1963, \textit{Asymptotic Behavior and Stability Problems in
Ordinary Differential Equations}, 2nd. edn., Springer-Verlag, Berlin

Chandrasekhar S., 1937, Nature, 139, 757

Combes, F., Wiklind, T., 1996, in \textit{Cold Gas at High Redshift}, eds.
M.Bremer, P. van der Werf, H. Rottgering, and C. Carilli, Kluwer: Dordrecht,
p. 215

Cowie L.L., Songalia A., 1995, ApJ, 453, 596

Damour, T., Gibbons G.W., Taylor J.H., 1988, Phys. Rev. Lett., 61, 1151

Damour T., Nordtvedt K., 1993, Phys. Rev., D\ 48, 3436

Damour T., Dyson F., 1997, Nucl. Phys. B, 480, 37

Damour T., Polyakov A.M., 1994, Nucl. Phys. B, 423, 596

Davidson W., Evans A.B., 1973, Int. J. Theo. Phys., 7, 353

Davidson W., Evans A.B., 1977, Comm. Roy. Soc. Edin. (Phys. Sci.), 10, 123

Dicke R.H., 1957, Rev. Mod. Phys. 29, 355

Dicke R.H., 1964, in\textit{\ Relativity, Groups and Topology}, eds. De Witt
C., De Witt B., pp. 163-313, Gordon \& Breach, New York

Dirac P.A.M., 1937a, Nature, 139, 323

Dirac P.A.M., 1937b, Nature, 139, 1001

Dirac P.A.M., 1938, Proc. Roy. Soc., A 165, 199

Dixit V.V., Sher, M., 1988, Phys.Rev.D, 37, 1097

Drinkwater M., Webb J., Barrow J.D., Flambaum V., 1997, MNRAS (in press)

Duval C., Gibbons G., Horv\'athy P., 1991, Phys. Rev., D 43, 3907

Dyson, F., 1971, Sci. American (Sept.) 225, 25

Eddington A.S., 1923,\textit{\ The Mathematical Theory of Relativity},
Cambridge UP, London.

Evans A.B., 1974, Nature, 252, 109

Evans A.B., 1978, MNRAS, 183, 727

Farhi E., Guth A., 1987, Phys. Lett. B 183, 149

Fischbach E., et al, 1986, Phys. Rev. Lett., 56, 3

Freund P.G.O., 1982, Nucl. Phys., B209, 146

Gamow G., 1967a, Phys. Rev. Lett., 19, 759

Gamow G., 1967b, Phys. Rev. Lett., 19, 913

Garc\'ia Bellido J., Linde A., Linde D., 1994, Phys. Rev., D 50, 730

Garc\'ia-Berro E., Hernanz M., Isern J., Mochkovitch R., 1995, MNRAS, 277,
801

Green M., Schwarz J.H., 1984, Phys. Lett., B149, 117

Gurevich L.E., Finkelstein A.M., Ruban V.A., 1973, Astrophys. Sp. Sci. 22,
231

Guth A.H., 1981, Phys. Rev., D23, 347

Harrison, E.R., 1995, QJRAS 36, 193

Hawking S.W., 1972 Commun. Math. Phys. 25, 167 (1972)

Hawking S.W. 1974 Nature \textbf{248}, 30

Heckmann O., Sch\"ucking E., 1955, Z. Astrophysik, 38, 95

Heckmann O., Sch\"ucking E., 1959, Handbuch der Physik 53, 489

Hellings R.W., 1984, in General Relativity and Gravitation, eds. Bertotti
B., de Felice F., Pascolini A., Reidel, Dordrecht

Holman R., Kolb E.W., Vadas S., Wang Y., 1991, Phys. Rev., D 43, 3833

Jordan P., 1938, Naturwiss., 26, 417

Jordan P., 1952, \textit{Schwerkraft und Weltall, Grundlagen der Theor.
Kosmologie}, Braunschweig, Vieweg and Sohn

Kang G, 1996, Phys. Rev. D 54, 7483

Kolb E.W., Perry M.J., Walker T.P., 1986, Phys. Rev., D 33, 869

Kothari D.S., 1938, Nature, 142, 354

Kolitch S.J., Eardley D.M., 1995, Ann. Phys. (NY), 241, 128

Krauss L.M., White M., 1992, Ap. J., 394, 385

La D., Steinhardt P.J., 1989, Phys. Rev. Lett., 62, 376

Linde, A., 1990, \textit{Inflation and Quantum Cosmology}, Academic Press, NY

Lorentz-Petzold D., 1984, Astrophys. Sp. Sci., 98, 249

Lynden-Bell D., 1982 Observatory 102, 86

Marciano W.J., 1984, Phys. Rev. Lett., 52, 489

Maurette M., 1972, Ann. Rev. Nuc. Part. Sci. 26, 319

Mathiazhagen C., Johri V.B., 1984, Class. Quantum Gravity, 1, L29

McVittie G.C., 1978, MNRAS, 183, 749

Meshcherskii I.V., 1893, Astron. Nach., 132, 129

Meshcherskii I.V., 1949, \textit{Works in the Mechanics of Bodies of
Variable Mass}, Moscow-Leningrad Publ. House, Moscow

Milgrom M., 1983, Ap. J., 270, 365

Milne E., McCrea W.H., 1934, Quart. J. Math., 5, 73

Nariai H., 1969, Prog. Theo. Phys. 42, 544

Narlikar J.V., 1963, MNRAS, 126, 203

Narlikar J.V., Kembhavi A.K., 1980, Fund. Cosmic Phys., 6, 1

Nordvedt K., 1968, Phys. Rev. D169, 1017

O'Hanlon J., Tupper B.O.J., 1970, Nuovo Cim. 137, 305

Pagel B.E.J., 1977, MNRAS 179, 81P

Pagel B.E.J., 1983, Phil. Trans. Roy. Soc. A, 310, 245

Prestage J.D., Tjoelker R.L., Maleki, L., 1995, Phys. Rev. Lett., 74, 3511

Reasenberg R.D., 1983, Phil. Trans. Roy. Soc. Lond., A310, 227

Savedoff M.P., Vila S., 1964, AJ 69, 242

Serna A., Alimi J.M., 1996, Meudon preprint 'Scalar Tensor Cosmological
Models'

Shapiro I., 1990, in\ \textit{General Relativity and Gravitation}, eds.
Ashby N., Bartlett D., Wyss W., Cambridge UP, Cambridge

Shikin I.S., 1971, Sov. Phys. JETP, 32, 101

Shikin I.S., 1972, Sov. Phys. JETP, 34, 236

Shlyakhter A.I., 1976, Nature 264, 340

Shlyakhter A.I., 1983, Direct test of the time-independence of fundamental
nuclear constants using the Oklo natural reactor, ATOMKI Report A/1

Smolin, L., 1992, Classical and Quantum Gravity 9, 173

Steinhardt P.J., Accetta F.S., 1990, Phys. Rev. Lett., 64, 2740

Teller E., 1948, Phys. Rev., 73, 801

Tubbs A.D., Wolfe A.M., 1980, ApJ, 236, L105

Varshalovich D. A., Potekhin A. Y., 1995, Space Science Review, 74, 259

Varshalovich D. A., Potekhin A. Y., 1996, Pis'ma Astron. Zh., 22, 3

Varshalovich D.A., Panchuk V.E., Ivanchik A.V., 1996, Astron. Lett., 22, 6

Vila S.C., 1976, ApJ., 206, 213

Vinti J.P., 1974, MNRAS, 169, 417

Wagoner R.V., 1970, Phys. Rev., D 1, 3209

Weinberg, S.,1972,\textit{\ Gravitation and Cosmology}, Wiley, NY

Weyl H., 1919, Ann. Physik, 59, 129

Wiklind T., Combes F., 1994, A\&A, 286, L9

Wiklind T., Combes F., 1996, A\&A, 315, 86

Will C.M., 1993,\textit{\ Theory and Experiment in Gravitational Physics},
2nd edn., Cambridge UP, Cambridge

Zwicky F., 1939, Phys. Rev., 55, 726

\end{document}